quant-ph/0504176
\documentclass[pra,12pt,eqsecnum]{revtex4}
\usepackage{epsf}
\usepackage{graphicx}
\usepackage [cp1251] {inputenc}
\usepackage [english] {babel}
\inputencoding {cp1251}

\begin{document}

\def\BE{\begin{equation}}
\def\EE{\end{equation}}
\def\BY{\begin{eqnarray}}
\def\BEA{\begin{eqnarray}}\def\EY{\end{eqnarray}}\def\EEA{\end{eqnarray}}
\def\q{\vec q}
\def\r{\vec r}
\def\L{\label}
\def\nn{\nonumber}
\def\({\left (}
\def\){\right)}
\def\[{\left [}
\def\]{\right]}
\def\o{\overline}
\def\BA{\begin{array}}
\def\EA{\end{array}}
 \def\ds{\displaystyle}

\title{Laser photon statistics in the feedback loop}
\author{T.~Golubeva and Yu.~Golubev}
\address{V. A. Fock Physics Institute, St.~Petersburg State University,\\
ul Ul'yanovskaya 1, 198504 St.~Petersburg, Stary Petershof, Russia}
\date{\today}

\begin{abstract}

 A mere correspondence between the electron statistics and the photon one vanishes in
the feedback loop (FBL). It means that the direct photodetection, supplying us
with the electron statistics, does not provide us with a wished information about
the laser photon statistics. For getting this information we should think up another
measurement procedure, and we in the article suggest applying the
three-level laser as a auxiliary measuring device. This laser has impressive property,
namely, its photon statistics survive information about the
initial photon statistics of the laser which excites coherently the three-level
medium. Thus, if we choose the laser in the FBL as exciting the three-level
laser, then we have an possibility to evaluate its initial photon statistics by
means of direct detecting the three-level laser emission. Finally, this
approach allows us to conclude the feedback is not capable of creating a
regularity in the laser light beam. Contrary, the final photon
fluctuations turn out to be always even bigger. The mentioned above feature of
the three-level laser takes place only for the strong interaction between the
lasers (exciting and excited). It means the initial state of the exciting laser
is changed dramatically, so our measurement procedure can not be identified with
some non-demolition one.

\end{abstract}

\maketitle

\section{Introduction}

After the very first works devoted to the laser sources of sub-Poissonian light
\cite{Golubev,Yammamoto}, the efforts to find physical systems able to emit an
effectively squeezed light and suitable for the aims of quantum optics have been
continued. There are a few ways to achieve the required light properties; each
of them has its own specific features. Here we are going to discuss only one of
the ways which is connected with the laser in the feedback loop (FBL) and is
likely the most problematic from the ideological point of view.

The idea of an experiment with a feedback is as follows. Laser emission is
detected by a photo-detector, and then the photocurrent is used for pump rate
correction of the laser medium in a manner such that, for example, positive
photocurrent fluctuation results in negative  fluctuation of pump rate of the
laser medium, and so it leads to negative fluctuation of operation power and,
accordingly, to negative fluctuation of the photocurrent. Thus, the photocurrent
occurs to be stabilized on specific temporal intervals. As was persuasively
shown in experiments \cite {Yammamoto, Fofanov, Mosalov}, it is possible to
reduce a photocurrent shot noise  even below the shot level in this way.

Unfortunately, it is impossible to say in this case that smoothing the fluctuations
in the photocurrent means corresponding smoothing in the photon fluctuations.
Really, the sub-shot noise in the electrical current is easy
understood for both the sub-Poissonian photon flux and the classical light beam
for which, certainly, we have no any reasons to say about the non-classical statistics.
It means in the FBL a mere correspondence between the electron statistics and photon
one vanishes, and we will say that  in this case a direct observation  of the photon
statistics is impossible.

To make the correct conclusion about the photon statistics in the FBL we should
provide some special measuring procedure. For example,
 any idea of quantum non-demolition procedure could be quite productive. However, now we
 do not see how it could be really organized.

At the same time, the quantum non-demolition procedure is not only what could be
offered. It is possible to imagine another measuring procedure in which  the
a state of the tested laser in the FBL  does not survive
(in contradiction to the quantum non-demolition one) but
the information about this state  is transferred to another element
('the measuring device'). If this device is
 outside the FBL, then
the information is available to be read under direct observation.
This gives us the possibility making
correct conclusion about the statistics of the laser emission in the FBL . In
this paper we will discuss one of the similar measurement schemes and offer to
test the three-level laser as a 'measuring device'.

Finally, we would like to mention the paper by Wiseman and Milburn
\cite {Milburn} which is devoted to the same problem. We believe, the main
conclusion there that the FBL is unable to lead to quantum effects in the laser
field is physically quite justified. At the same time, in our opinion, the
theoretical base of  the discussion in the work  does not seem
to us quite convincing. The models used there are not always adequate to
investigated processes. And, what is more important, authors do not offer any
measuring procedure, and make the conclusions at the mathematical level, by
analyzing the equations for intracavity field.

The organization of our paper is as follows. In sec.~\ref{II}, the quantum
theory of the simplest single-mode laser in the form of  Langevin's equations
for the photon number is represented. In sec.~\ref{III}, the phenomenological
models of photodetecting the laser emission and the negative feedback are
discussed. In sec.~\ref{IV}, the Langevin theory for two jointly working lasers
is represented. It is proved that under the conditions of the strong coupling
between the lasers, statistical properties of the three-level laser emission keep
the information about the photon statistics of the coherently exciting laser.
In sec.~\ref{V}, we choose the laser in the FBL as the coherently exciting the
three-level lasing. We make the conclusion about the photon statistics in the
emission of the laser in the FBL by means of analyzing emission from the
three-level laser.

\section{The quantum Langevin theory of single-mode laser}\L{II}
In this section we shall give a brief resume of the quantum laser model
developed in Ref.~\cite{Benkert,Kolobov}. We shall define the physical
parameters of this model and get the equations which will be used in the
following sections.

In Fig.~1 we have shown  the pump process to the upper laser state
$|1\rangle$ with mean rate $R$. For stationary in time average pumping rate, an
influence of the pump statistics can be characterized by the single parameter
$p\leq1$. For $p=1$ the pump is perfectly regular while for $p=0$ the pump has
Poissonian statistics. Intermediate values of $0\leq p\leq 1$ correspond to
sub-Poissonian pumping while for $p\leq 0$ the pump process possess the excess
classical fluctuations and corresponds to super-Poissonian statistics.

 \begin{figure}
  \includegraphics[width=50mm]{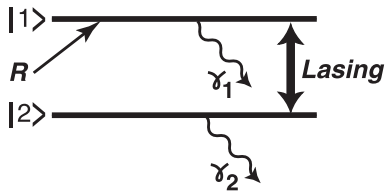}
 \caption{Atomic energetic configuration for the two-level laser}
 \label{fig1}
 \end{figure}

This pump statistics was introduced into the quantum laser model using the
Heisenberg-Langevin equations for the operator-valued collective populations  of
the upper and lower levels (see Fig.~1), and for the collective polarization.  At
the same time, there is a way to reformulate the Heisenberg-Langevin theory via
the c-number values and derive so-called  the Langevin theory. Here we will
apply  the latter treatment, the corresponding equations read \cite{Benkert,Kolobov}:
\BY
&&\dot \alpha=-\frac{\kappa}{2}\alpha+gP,\L{2.1}\\
&&\dot P=-\gamma_\perp P+g(N_1-N_2)\alpha+F_p,\\
&&\dot N_1=R-\gamma_1 N_1-g(\alpha^\ast P+\alpha P^\ast)+F_1,\\
&&\dot N_2=-\gamma_2 N_2+g(\alpha^\ast P+\alpha P^\ast)+F_2.\L{2.4}
\EY
Here $\alpha$ is the complex field (mode) amplitude, $P$ is the complex collective
polarization of the laser transition, $N_1(N_2)$ is the collective population of
the upper (lower) laser level. $R$ is the mean rate of pump to the upper laser
level, and $\kappa$ is the spectral width of the laser mode,
$\gamma_{1,2}$ and $\gamma_\perp$ are the corresponding atomic longitudinal and transverse
relaxation constants and
$g$ is the coupling constant ensuring the dipole atom-laser field interaction.

$F_1,\;F_2,\;F_P$ are the stochastic sources which are specified in the work
\cite{Benkert,Kolobov} in the general case. By taking away the sources, we get
the corresponding semiclassical theory. Because in the stationary regime the
usual laser conditions guarantee relatively small fluctuations of the
populations and the polarization, we have a right to hold that the semiclassical
solutions coincide with the mean values with high precision. Putting a
simplified conditions $\gamma_1\ll\gamma_2$ and
$2\gamma_\perp=\gamma_1+\gamma_2\approx\gamma_2$, it is easy
 to obtain the following stationary semiclassical
solutions:
\BY
&&g\;\o{\alpha^\ast P}=\frac{R}{2}\frac{I}{1+I},\qquad \gamma_1\o
N_1=R\;\frac{1}{1+I},\qquad \gamma_2\o N_2=R\;\frac{I}{1+I}.
\EY
Here the dimensionless power of lasing is defined as $I=\beta
n\;\;(n=\o{|\alpha|^2})$, and
\BY
\beta^{-1}=\gamma_\perp\gamma_1/(2g^2)
\EY
is the photon number saturating the laser transition. Taking
into account that under the stationary generation
\BY
&& 2g\;\o{\alpha^\ast P}=\kappa
\EY
and, as a result, $R=\kappa/\beta (1+I)$, the same solutions can
be rewritten in the form
\BY
\gamma_1\o N_1=\kappa/\beta,\qquad\gamma_2\o N_2=\kappa n
\;\;(\o N_2=0).
\EY
To calculate the quantum fluctuations $\delta N_{1,2}=N_{1,2}-\o N_{1,2}$,
$\delta P=P-\o P$ and $\delta \alpha=\alpha-\o \alpha$ around the stationary
solutions we shall linearize Eqs.~(\ref{2.1})-(\ref{2.4}) relative to these
fluctuations (neglecting by the phase diffusion). In particular, in this approximation, the non-zero correlation
functions for the stochastic sources are:
 \BY
&& \o{F_1(t)F_1(t^\prime)}=\kappa/\beta
\[2-p(1+I)\]\;\delta(t-t^\prime),\\
&& \o{F_1(t)F_2(t^\prime)}=\kappa n\;\delta(t-t^\prime),\\
&& \o{F_p^\ast(t)F_p(t^\prime)}=\kappa/\beta\;
\gamma_2/\gamma_1\;\delta(t-t^\prime),\\
&& \o{F_p(t)F_p(t^\prime)}=\kappa\o{\alpha}^2\;\delta(t-t^\prime),\\
&& \o{F_p(t)F_2(t^\prime)}=\kappa\gamma_2/(2g)\;\o\alpha\;\delta(t-t^\prime).
\EY
To specify more our physical conditions let us choose
$\kappa\ll\gamma_{1}$ (the high-Q cavity approximation). It
means the field amplitude $\alpha $ is developed more slowly
than the atomic variables and we have a right to apply the
adiabatical approximation putting in our theory $\dot P=0$ and $
\dot N_{1,2}=0$. It allows us to derive  the single equation for
the complex field amplitude $\alpha=\sqrt u\exp(i\varphi)$ or
even for the photon number $u$:
\BY
&&\dot u=-\kappa u +R+F,\qquad \o{F(t)F(t^\prime)}=-p\kappa
n\;\delta(t-t^\prime).\L{2.14}
\EY
Here we have selected more interesting case $I\gg1$ (the saturation regime).

It needs to stress, the equation (\ref{2.14}) could be easy
obtained within the consideration in the work \cite{Golubev},
although the master equation approach was developed there.

\section{Formal models of photodetecting the laser emission
and the negative feedback}\L{III}
In optical experiments we follow the photocurrent, and according to the
photoeffect law its mean value is equal to a mean light stream falling on the
photodetector, i.e., in the case of the single mode laser $ \o i=\kappa n,\;n= \o u$. To
write the corresponding equality without averaging it is not enough simply taking
the symbol of averaging away. We have to take into account that
the process of the atomic ionization under photodetecting
 is essentially random and then to write the random photocurrent
 phenomenologically in the form:
\BY
i(t)=\kappa u(t)+S(t),\qquad\o S=0,\qquad\o{S( 0)S(t)}=\o i\;\delta(t)\L{2.15}.
\EY
Here the stochastic source $S(t)$ ensures the appearance in the
photocurrent of the so-called shot noise.

Let us rewrite the equations (\ref {2.14}), (\ref {2.15}) in the
spectral representation:
 \BY
-\!\!\!\!&&i\omega\varepsilon_\omega=\kappa\varepsilon_\omega+
F_\omega\L{3.4},\\
&&\delta i_\omega=\kappa \varepsilon_\omega+S_\omega,\L{3.5}
\EY
where
 \BY
&& \varepsilon_\omega=\frac{1}{\sqrt{2\pi}}\int\limits_{-\infty}^{+\infty}dt\;
e^{i\omega t}[u(t)- n],\qquad  \delta i_\omega=\frac{1}{\sqrt{2\pi}}\int\limits_{-\infty}^{+\infty}dt\;
e^{i\omega t}[i(t)-\o i]
\EY
are the Fourier components respectively for the photon number and photocurrent
fluctuations. The spectral components $F_\omega$ and $S_\omega$ are written in
the same way. The non-zero correlation functions for them are:
\BY
&&\o{F_\omega F_{\omega^\prime}}=-p\;\o i\;\delta(\omega+\omega^\prime)\\
&& \o{S_\omega S_{\omega^\prime}}=\o i\;\delta(\omega+\omega^\prime)
\EY
Often investigators follow the spectral density of the photocurrent
fluctuations $ (\delta i^2) _ \omega $ (the photocurrent spectrum) which is
introduced by the correlation function:
\BY
\o{\delta i_\omega\;\delta i_{\omega^\prime}}=(\delta
i^2)_\omega\;\delta(\omega+\omega^\prime).
\EY
Now it is not difficult to get with help of the formulas in this section that
\BY
&&(\delta i^2)_\omega/\o i=1-p\kappa^2/(\kappa^2+\omega^2).\L{3.9}
\EY
One can see, for the random pump of the laser medium $p=0$ the current
spectrum contains only the shot noise term. At the same, for the regular pump $p=1$
the shot noise turns out to be reduced on the zero frequency. By this formula
for the photocurrent we have a right to make a correct conclusion relative
to the light flux.  The Poissonian photon statistics take
place in the case of $p=0$ and,  the sub-Poissonian ones - in the case of $p=1$.

For the formal introduction of the feedback into the equations a
merely phenomenological approach is usually used; with this
approach, the mean rate of the laser pump of medium $R $ in the
equation (\ref {2.1}) is changed so that it is decreased with
the photocurrent fluctuation increasing $ \delta i (t) =i (t)-\o
i $, for example, by the rule:
\BY
r\to r\(1-\lambda\;\frac{\delta i}{\o i}\),\L{2.2}
\EY
where the parameter $ \lambda $ could be called as a feedback efficiency.
Certainly, it is primitive model of the FBL but it is enough for our
reasons.

Applying the replacement (\ref{2.2}) in initial laser equations
(\ref{2.1})-(\ref{2.4}), we are able to get the equations for the fluctuations
in full analogous with equations (\ref{3.4})-(\ref{3.5}) which read
 \BY
&&\[(1+\lambda)\kappa-i\omega\]\varepsilon_\omega= F_\omega-\lambda
S_\omega\L{11}\\
&&\delta i_\omega=\kappa \varepsilon_\omega+S_\omega\L{12}.
\EY
The algebraic equations allow us to calculate the photocurrent spectrum which is
given by
\BY
&&(\delta i^2)_\omega/\o i=1-
\frac{p-1+(1+\lambda)^2}{(1+\lambda)^2+\omega^2/\kappa^2}.\L{}
\EY
It is readily seen that if the FBL efficiency  is high enough $(\lambda\gg1)$, then the
complete reduction of the shot noise in the photocurrent on the zero frequency
 takes  place
independently of the pump statistics. However, we shall remember that this
conclusion can not be extended to the photon statistics. Our aim here is to
study the photon statistics with help of the auxiliary measuring  procedure.

\section{Strong coupling between exciting and
excited lasers } \L{IV}
As mentioned above,  the three-level lasing with the coherent
excitation by the emission from another laser has extremely
attractive feature. Under the specific conditions, the
statistical features in its radiation replicate the statistical
features of the exciting radiation. However, it takes place only
when the interaction between lasers is essentially strong. This
idea can be explained with the help of the mental experiment
which is represented in Fig.\ 2. One can see, there are a
two-level exciting laser (the '2-laser') and a three-level
excited one (the '3-laser'). They occupy a position such that
they have the common intracavity space. It is assumed that the
three-level medium is coherently excited by the intracavity
field of the two-level laser. And we assume that the intracavity
lifetime of exciting photon is much less than the lifetime in
absence of the three-level medium. Further, we will apply the
term 'strong coupling' between lasers for just this phenomenon.

 \begin{figure}
  \includegraphics[width=120mm]{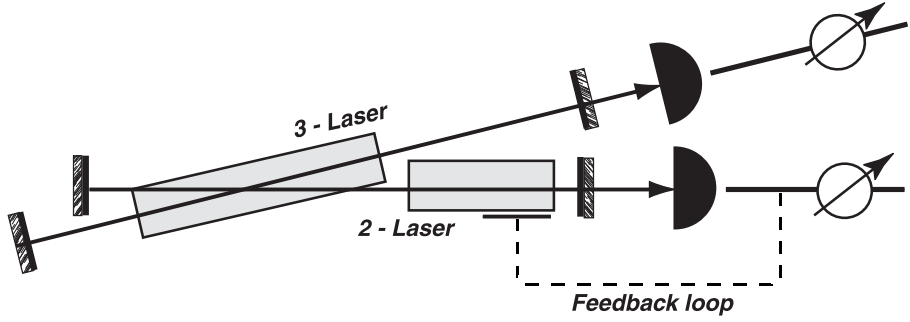}
 \caption{Mental experimental setup with two strongly coupled lasers.
 '2-laser' - the two-level exciting laser; '3-laser' - the three-level excited laser }
 \label{fig2}
 \end{figure}

In Fig. \ 3, both the two-level and three-level atomic
configurations are shown. As for the two-level medium (on the
left in the figure), as before, there are an incoherent pump
(regular or random) to the upper laser state $|1\rangle$ with
the mean rate $R$ and the spontaneous emissions with the rates
$\gamma_1$ and $\gamma_2$.

 \begin{figure}
  \includegraphics[width=120mm]{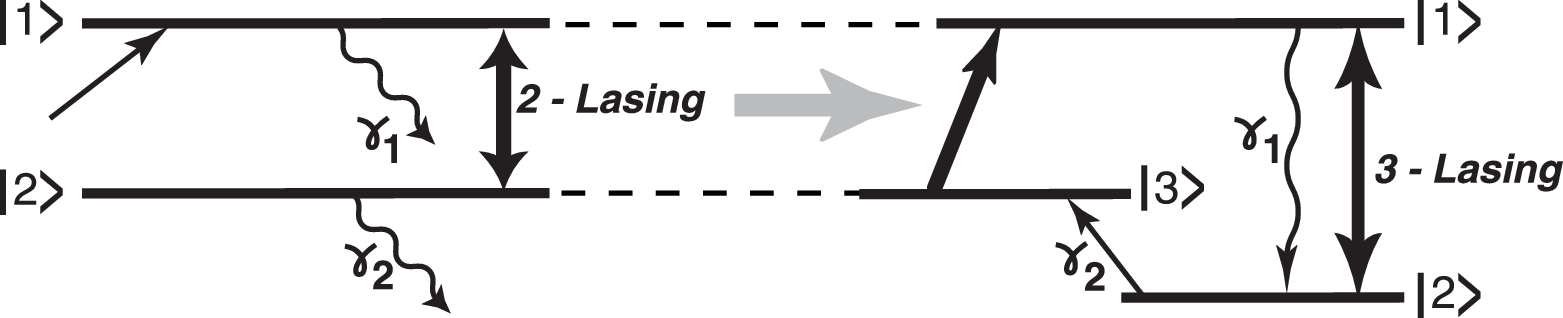}
 \caption{Atomic energetic configurations for 2-laser (on the left) and 3-laser
 (on the right) }
 \label{fig3}
 \end{figure}

In the three-level medium (on the right in the figure), there is
an incoherent pump with the rate $\tilde\gamma_2$. This ensures
populating the intermediate non-laser atomic state
$|\tilde3\rangle$. Take into account, hereinafter, the sign
'tilde' over any symbol tells us that this symbol is concerned
the three-level laser.

It is suggested that the three-level oscillation takes place on
the transition $|\tilde1\rangle\leftrightarrow|\tilde2\rangle$
and this is realized by a joint action of both the incoherent
$|\tilde2\rangle\to|\tilde3\rangle$ and coherent
$|\tilde3\rangle\to|\tilde1\rangle$ pumps. Our two-level laser
ensures the last.

Besides, the spontaneous emission on the laser transition with
the rate $ \tilde\gamma_1$ is taken into account.

We shall construct the quantum theory of two interacting lasers by applying the
Langevin treatment as before.
 Certainly, a structure of the equations is  more
complicated. Really, now we have to discuss the populations of five atomic states
(two-level and three-level medium) and the polarizations of four atomic transitions
instead of two populations and one polarization.

Besides, we have to introduce two laser amplitudes $\alpha=\sqrt
u\exp(i\varphi)$ for the exciting laser  and $\tilde\alpha=\sqrt
{\tilde u}\exp(i\tilde\varphi)$ for the excited one
 instead of one $\alpha$  and try
to construct the equations for their fluctuations
$\varepsilon=u-n$ and $\tilde\varepsilon=\tilde u-\tilde n$.
This analysis is relatively easy carried out and the following
system of the equations in the spectral representation is
obtained:
 \BY
&&-i\omega\varepsilon_\omega=-(\kappa+\kappa_0)\varepsilon_\omega
+\tilde\kappa \tilde \varepsilon_{\omega}+F_\omega,\L{4.1}\\
&&-i\omega\tilde\varepsilon_{\omega}=-2\tilde\kappa\tilde\varepsilon_{\omega}
+\kappa_0\varepsilon_\omega+\tilde F_{\omega}\L{4.2}.
\EY
Here  $\kappa,\tilde \kappa$ are the spectral mode widths;
$n,\;\tilde n$ are the corresponding mean photon numbers under
the stationary oscillation. One can easily get some useful
relationships for the semiclassical values:
 \BY
\tilde n\tilde\kappa=n\kappa_0,\qquad(\kappa+\kappa_0)n=R.
\EY
The value
\BY
\kappa_0=\tilde\gamma_2\;\(\tilde g_{13}/\tilde
g_{12}\)^2\;\tilde N/\tilde n
\EY
is the rate  of the coherent pump of the three-level medium by
the field of two-level generation. In the formula $\tilde g_
{13},\tilde g_{12} $ are the constants of the dipole interaction
between the field and the three-level medium on the transitions
$ (1-3) $ and $ (1-2) $, accordingly; $\tilde N $ is the number
of the three-level atoms participating in laser process.

The stochastic sources are specified by the correlation functions
 \BY
&&\o{F_\omega F_{\omega^\prime}}=-p(\kappa+\kappa_0) n\;
\delta(\omega+\omega^\prime),\L{3.2}\\
&&\o{\tilde F_{\omega}\tilde
F_{\omega^\prime}}=-2\tilde\kappa\tilde n
\;\delta(\omega+\omega^\prime).
\EY
Note that the theory of two lasers operating jointly have been
considered in Ref.~\cite{Rost} but in the master equation
treatment. Although we have written
 the equations
(\ref{4.1})-(\ref{4.2})  on the basis of the Langevin approach,
nevertheless they could easily be  obtained from the cited work
on the base of the master equation.

As before, we should derive the phenomenological expressions for the photocurrent
fluctuations:
\BY
 &&\delta i_\omega=\kappa\varepsilon_\omega+S_\omega,
 \qquad\o{S_\omega S_{\omega^\prime}}=\o i
\;\delta(\omega+\omega^\prime),\L{4.7}\\
&&\delta\tilde
i_{\omega}=\tilde\kappa\tilde\varepsilon_{\omega}+\tilde
S_{\omega}, \qquad\o{\tilde S_{\omega}\tilde
S_{\omega^\prime}}=\o{\tilde i}
\;\delta(\omega+\omega^\prime)\L{4.8}.
\EY
The equations are derived on the assumption that there are no any nonlinear effects
at the interaction of the two-level generation with the three-level medium; it
means that we keep only the main term in the perturbation theory concerning
to the coherent excitation of the three-level medium.
 The next
assumption  $\tilde \gamma_{2} \ll\tilde\gamma_{1}$ is made here
only for simplification.

The formulas above allow us to derive the photocurrent spectrum
under the registration of the emission of the three-level laser.
It is given by the formula:
\BY
&&(\delta \tilde i^2)_\omega/\o {\tilde
i}=1-2\tilde\kappa^2\frac{\omega^2+\kappa(\kappa+\kappa_0)
+\kappa_0(\kappa+\kappa_0)p/2}{\(\omega^2-\tilde\kappa(2\kappa+\kappa_0)
\)^2+\omega^2\(2\tilde\kappa+\kappa+\kappa_0\)^2 }.\L{4.9}
\EY
It is not difficult to obtain that this coincides exactly with
(\ref{3.9}) as $\kappa=\tilde\kappa\ll\kappa_0$. So if the
photocurrent spectrum under registration of the laser emission
is known and we choose this laser for coherent excitation (under
strong coupling) of the three-level lasing, then the
photocurrent spectrum from the three-level laser is the same. It
means one can study the photocurrent spectrum from the exciting
laser without the three-level medium by means of applying the
measuring device, i.e., by  studying the emission from the
three-level laser.

\section{The laser emission statistics in the FBL}\L{V}
The results of the previous section allows us to consider the three-level laser
as the 'measuring device' which is reasonable for applying when the statistical
analysis of light under the direct photodetection is impossible
for any reason (for example,
in the case of the laser in the FBL).

Thus in this section we choose the laser in the FBL as exciting
one for the three-level laser.  To rewrite equations (\ref
{4.1}) - (\ref {4.2}) for the case of a laser in the FBL we have
to make the replacement (\ref {2.2}) in the initial equations.
Then instead of equations (\ref {4.1}) - (\ref {4.2}) we can
get:
 \BY
&&\[(\kappa+\kappa_0)(1+\lambda)-i\omega\]\varepsilon_\omega=\tilde\kappa
\tilde\varepsilon_{\omega}+F_\omega-\lambda(1+\kappa_0/\kappa)S_\omega,\L{5.1}\\
&&(2\tilde\kappa-i\omega)\tilde\varepsilon_{\omega}=\kappa_0\varepsilon_\omega+\tilde
F_{\omega}\L{5.2}.
\EY
These algebraic equations together with (\ref {4.7}) -
(\ref{4.8}) allow us to obtain the photocurrent spectrum
$(\delta\tilde i^2)_\omega$ in the explicit form:
\BY
&&(\delta\tilde i^2)_\omega/\o{\tilde i}=1-\\
&&- 2\tilde\kappa^2
\;\frac{\omega^2+\kappa^2(1+x)\[1+\lambda(2+x)
+\lambda^2(1+x)(1-x/2)\]}{
\[\omega^2-\tilde\kappa\kappa\(2+x+2\lambda(1+x)\)\]^2
+\omega^2\[2\tilde\kappa+\kappa(1+x)(1+\lambda)\]^2},\qquad
x=\frac{\kappa_0}{\kappa}\L{5.3}.\nn
 \EY
 This result is written for case the random pump in the two-level laser.
 Just this case is of our interest because we would like to understand
 whether the quantum effects appear  in the
FBL if they were absent before.

 Now
we have a possibility to realize what happens with arising the
FBL. Let the feedback efficiency be high enough $\lambda\gg1$,
and there be a strong coupling $\kappa=\tilde\kappa\ll\kappa_0$,
then
\BY
&&(\delta \tilde i^2)_\omega/\o{\tilde
i}=1+\frac{\kappa_0}{4\kappa}\;
\frac{4\kappa^2}{\omega^2+4\kappa^2}.
 \EY
In accordance with our logic in this work and with the last
formula, we have a right to say that the noises on the zero
frequency  of the laser emission in the FBL are essentially
above the shot level because it is the case for the emission of
our measuring device. Thus, although the feedback causes the
sub-Poissonian statistics for the photoelectrons, nevertheless
the photon flux in the FBL turns out to be super-Poissonian.

This work was performed within the Franco-Russian cooperation program "Lasers
and Advanced Optical Information Technologies" with financial support from the
following organizations: INTAS (Grant No. INTAS-01-2097), RFBR (Grant No.
03-02-16035).

 \begin{figure}[t]
  \includegraphics[width=120mm]{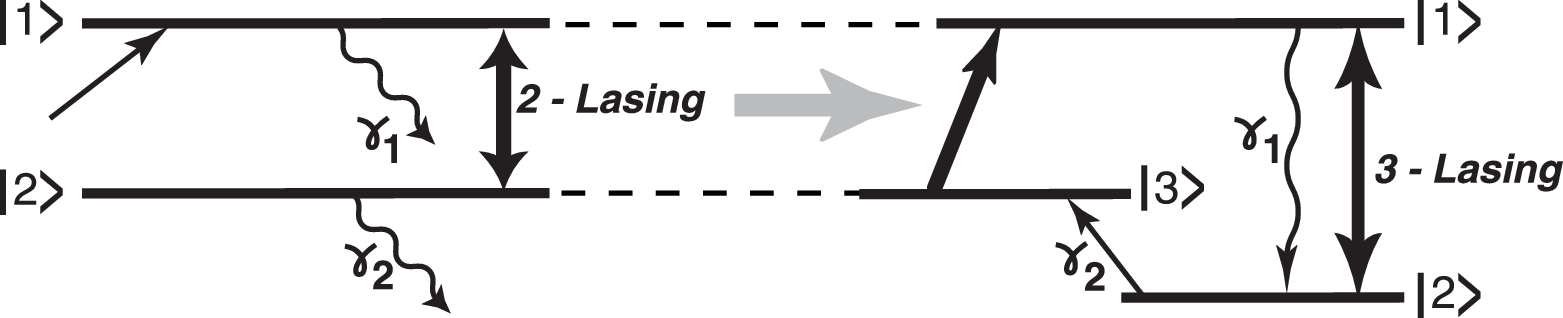}
 \caption{Atomic energetic configurations for 2-laser (on the left) and 3-laser
 (on the right) }
 \label{fig3}
 \end{figure}

\end{document}